\documentclass{raa01}            

\usepackage{natbib}
\usepackage{url}
\bibpunct{(}{)}{;}{a}{}{,}
\usepackage{graphicx,times}             

\begin{document}

   \title{Stellar populations in star clusters}

   \volnopage{{\bf} Vol.\ {\bf} No. {\bf},
 {\small}}

   \author{Cheng-Yuan Li
      \inst{1,2}
     \and Richard de Grijs
      \inst{3,4}
     \and Li-Cai Deng
      \inst{5}
   }

   \institute{Department of Physics and Astronomy, Macquarie University, North Ryde, NSW 2109, Australia; {\it chengyuan.li@mq.edu.au} \\
        \and
              Purple Mountain Observatory, Chinese Academy of Sciences, Nanjing 210008, China
        \and
             Kavli Institute for Astronomy \& Astrophysics and Department of Astronomy, Peking University, Yi He Yuan Lu 5, Hai
  Dian District, Beijing 100871, China
        \and
             International Space Science Institute--Beijing, 1
  Nanertiao, Zhongguancun, Hai Dian District, Beijing 100190, China
        \and
             Key Laboratory for Optical Astronomy, National
  Astronomical Observatories, Chinese Academy of Sciences, 20A Datun
  Road, Chaoyang District, Beijing 100012, China\\
\vs\no
   {\small}}

\abstract{Stellar populations contain the most important information
  about star cluster formation and evolution. Until several decades
  ago, star clusters were believed to be ideal laboratories for
  studies of simple stellar populations (SSPs). However, discoveries
  of multiple stellar populations in Galactic globular clusters have
  expanded our view on stellar populations in star clusters. They have
  simultaneously generated a number of controversies, particularly as
  to whether young star clusters may have the same origin as old
  globular clusters. In addition, extensive studies have revealed that
  the SSP scenario does not seem to hold for some intermediate-age and
  young star clusters either, thus making the origin of multiple
  stellar populations in star clusters even more complicated. Stellar
  population anomalies in numerous star clusters are well-documented,
  implying that the notion of star clusters as true SSPs faces serious
  challenges. In this review, we focus on stellar populations in
  massive clusters with different ages. We present the history and
  progress of research in this active field, as well as some of the
  most recent improvements, including observational results and
  scenarios that have been proposed to explain the
  observations. Although our current ability to determine the origin
  of multiple stellar populations in star clusters is unsatisfactory,
  we propose a number of promising projects that may contribute to a
  significantly improved understanding of this subject.}


\authorrunning{C.-Y Li et al.}
\titlerunning{Stellar populations in star clusters}

   \maketitle
\section{Introduction}           
\label{S1}

Star clusters are the basic units of star formation \citep{Lada03a}:
almost all stars form in clustered environments. Current consensus on
the formation of star clusters suggests that most stars form tracing
the turbulent structure of the interstellar medium and in an initially
supervirial state. Within an extremely short period (about one
crossing time), the initially turbulent, `fractal' structures will
collapse into bound clusters
\citep{Bonn08a,Alli09a,Giri12a,Moec15a}. Subsequently, a large
proportion will gradually dissipate into the galactic field
\citep{grijs10a}. Understanding the stellar populations of star
clusters is, therefore, of fundamental importance for understanding
many astrophysical processes, including star cluster formation and
evolution, the chemical evolution of Galactic stellar populations, as
well as the stellar dynamics in star clusters.

The `simple stellar population' (SSP) scenario is the assumption that
stars in a star cluster all originate from a common progenitor giant
molecular cloud (GMC), during the same era, and thus they would share
similar metallicities. It has been confirmed that the initial
star-forming process in star clusters approximately resembles a single
burst \citep{Cabr14a}. The combination of gas expulsion owing to
energetic photons ejected by the most massive first-generation stars
and the strong stellar winds triggered by the first batch of Type II
supernovae will quickly exhaust all of the gas in the GMC, thus
quenching the star- and cluster-forming process \citep{Bast14a}. The
nature of most open clusters (OCs) and young massive clusters (YMCs)
has been confirmed as resembling SSPs.

During the last few decades, observations have revealed the presence
of multiple stellar populations in Galactic globular clusters
(GCs). The observational evidence can be classified into photometric
and spectroscopic evidence. The former refers to the fact that the
photometric color--magnitude diagrams (CMDs) of some GCs display
multiple distinct features in or along their main sequences
\citep[MSs; e.g., NGC 2808:][]{Piot07a}, their subgiant branches
\citep[SGBs; e.g., 47 Tuc:][]{Ande09a}, their red-giant branches
\citep[RGBs; e.g., NGC 288:][]{Piot13a}, or even in their horizontal
branches \citep[HBs; e.g., NGC 2808:][]{Bedi00a}. Sometimes, individual
GCs even display a combination of such multiple features. In the
photometric CMDs, these features can be explained as the result of a
diversity of ages, helium abundances, and metallicities. Since SSP
stars formed at approximately the same time from a common molecular
cloud, variations in age, helium, and metal abundance hence
unambiguously reflect the occurrence of more than a single
star-forming episode during their host clusters' evolution.

GCs have also been subject to intense scrutiny based on spectroscopic
analyses, which have revealed star-to-star chemical dispersions of
specific elements. \cite{Cohe78a} first noted that the Na abundance
variations of RGB stars in the GCs M3 and M13 exceeded the
observational errors. Her pioneering work has stood the test of time:
subsequent measurements of Galactic GCs have uncovered the well-known
Na--O anticorrelation, where oxygen-depleted stars have higher sodium
abundance \citep{Carr04a,Grat04a,Carr09a}. Another well-studied
Galactic GC relationship is the Mg--Al anticorrelation, which is most
easily seen in intermediate-metallicity clusters \citep[e.g.,
  M13;][]{Shet96a}. The large scatter in the abundances of some
specific elements is not expected if these GCs were SSPs. A
straightforward explanation of the scatter properties is that there
has been secondary star formation, fueled by abundance-enhanced
material. In this review we will not discuss the chemical evolution of
stellar populations in depth, since this aspect was introduced in
great detail in the review of \cite{Grat04a}.

A key problem associated with the stellar populations in star clusters
relates to an important open question: do GCs have the same origin as
OCs and YMCs? It seems that the latter, very young objects may
encounter significant difficulties to survive to the cosmic age
($\geq$ 10 Gyr) owing to a range of effects, including internal
two-body relaxation, which causes most stars to `evaporate' from the
cluster \citep{Spit87a}. Only those clusters that initially contained
at least 10$^5$ stars can avoid disruption within 10 Gyr
\citep{Port10a}. On the other hand, in terms of their initial masses,
only young star clusters with masses between $10^5$ $M_{\odot}$ and
$10^6$ $M_{\odot}$ have the capacity to capture their initial runaway
gas. Because the subsequent Type II supernovae explosions would
further accelerate the runway gas flows, the mass threshold for a YMC
to capture its initial runaway gas would increase dramatically. If a
star cluster can not marshall additional gas reserves, its
star-forming process will cease rapidly. As regards the stellar
populations in star clusters, almost all observed OCs and YMCs will
fail to generate multiple stellar populations, which, however, is a
common feature of most observed GCs \citep[see][his Chapter
  4]{LiTh_15}.

This review is organized as follows. In Section \ref{S2} we show that
the SSP approximation is not a tentative assumption but based on
convincing evidence. Readers will realize why the presence of multiple
stellar populations in GCs (and also in some young and
intermediate-age clusters) is a problem. In Section \ref{S3} we
discuss observational results of stellar populations in star clusters
with different ages. In Section \ref{S4} we compare and contrast
currently popular scenarios that aim at explaining the origin of
multiple stellar populations; their advantages and disadvantages are
also discussed. We next discuss, in Section \ref{S5}, how one can go
about examining some of these compelling and controversial
scenarios. A brief summary is given in the concluding section (Section
\ref{S6}).

\section{Thoughts on Stellar Populations in Star Clusters}\label{S2}

The first issue confronting any fundamental discussion of stellar
populations is the star-formation mode in star
clusters. Theoretically, except for the most massive and most compact
star clusters, stars in a star cluster should form in single-burst
mode: the star-formation process in a star cluster will cease rapidly
after formation of the first-generation stars owing to the quick
exhaustion of the initial gas, which is mainly caused by stellar
wind-induced mass loss \citep[e.g., energetic photons ejected by the
  most massive stars;][]{Long14a} and supernova explosions. The
typical escape velocity should be comparable to the sound speed in
ionized hydrogen (H{\sc ii}) gas \citep[$\sim$ 10 km
  s$^{-1}$;][]{Bane15a}. For some massive clusters, gas expulsion may
be dominated by radiation pressure, and therefore their escape
velocities could even exceed 10 km s$^{-1}$ \citep{Krum09a}. The
timescale of gas expulsion driven by these initial stellar winds is
very short, usually less than or comparable to the cluster's crossing
time ($\sim$ 1 Myr). If a single supernova explosion from the rapid
evolution of the most massive star occurs at this early stage, this
timescale may even be close to zero, i.e., a cluster will be exposed
(devoid of gas) almost immediately after birth.

Even if a cluster can retain its gas until the end of the initial gas
expulsion phase ($\leq$ 1 Myr), during the period from 3 Myr to 10 Myr
most O-type stars ($\sim 16 M_{\odot}$ to 25 $M_{\odot}$) evolve off
the MS. A 15 $M_{\odot}$ star could eject 3$\times$10$^{50}$ erg into
the interstellar medium over the period of 0.1 Myr, which is
sufficient to unbind 10$^4$ $M_{\odot}$ gas within a 1 pc region
\citep{Baum07a}. Therefore, a large fraction of residual gas in a
cluster with an age between 3 Myr an 10 Myr is not expected. During
this phase, Type II supernova explosions will occur frequently. These
multiple supernova explosions will be the `death blow' to any residual
gas remaining in the cluster. Supernova explosions will expel all
stellar material at a velocity of up to $3\times 10^4$ km s$^{-1}$
(10\% of the speed of light), driving a shock wave through the
interstellar medium of their host star clusters. Such a shock wave
will accelerate all remaining gas to a velocity of several 100 km
s$^{-1}$. Calculations have shown that only clusters that have formed
from the initial condensations with masses of 10$^7$ $M_{\odot}$ to
10$^8$ $M_{\odot}$ can survive the first series of supernova
explosions \citep{Shuz00a}.

Because the initial star-formation process ceases very quickly, the
age range of the first-generation stars would thus be constrained to a
timescale of several million years.  This timescale, compared with the
typical ages of most GCs ($\sim$ 10 Gyr) and YMCs (10 Myr -- 100 Myr),
is indeed very short.

Stars that are less massive than $\sim$ 8 $M_{\odot}$ will undergo the
post-RGB and asymptotic giant branch (AGB) phases, when they 
evolve to their final evolutionary stages. The ages of their host star 
clusters must be {\it at least} 30 Myr, which is already equivalent 
to the typical age of most YMCs. These intermediate-mass 
AGB stars will deposit most of their stellar material into the 
interstellar medium, which is thought to be important for 
secondary star formation: ejected stellar material from
AGB stellar winds forms the building materials of newly born stars,
possibly leading to the formation of multiple stellar populations with
enhanced chemical abundances, at least compared with the initial
stellar generation in the cluster. Many models aim at explaining the
observed multiple-stellar populations in GCs are based on this type of
`self-enrichment' scenario \citep[i.e.,][]{Vent09a,Valc11a}. However,
since in most GCs the observed secondary stellar generation has a
comparable total mass to the first stellar generation, to explain the
observed high fractions of secondary stellar generations,
these scenarios have to assume extremely high cluster masses at early
times (about 10 to 100 times their current masses). In addition, the
velocity of the AGB stellar winds is comparable to the initial O-type
stars' stellar winds, which are $\sim$10 km s$^{-1}$ to 100 km
s$^{-1}$. Because initial gas expulsion will cause expansion of the
cluster to a less compact configuration and leads to the loss of a
fraction of the initial cluster mass, an exposed cluster is unlikely
able to retain these intermediate-mass AGB stellar winds.

Mass loss caused by evaporation will further complicate the estimates
\citep{McLa08a}. The amount of mass lost through this mechanism can be
calculated as
\begin{equation}
\dot{m} \approx 1100\left(\frac{\rho_{\rm h}}{M_{\odot}\,{\rm
    pc}^{-3}}\right)^{-1/2}\, {M_{\odot}}\,{\rm Gyr}^{-1},
\end{equation}
where $\rho_{\rm h}$ is the gas density inside the cluster's half-mass
radius, $r_{\rm h}$. Assuming a typical GC age of 12 Gyr, most GCs
should initially have been 10 to 15 times more massive than their
current mass, indicating a strong capacity to retain their initial
runaway gas. However, for OCs and YMCs, most of which are younger than
$\sim$100 Myr, the expected amount of mass loss is much
smaller. Therefore, their initial masses do not reach the mass
threshold for capturing runaway gas.

\begin{figure}[htbp]
\centering
\includegraphics[width=4.5in]{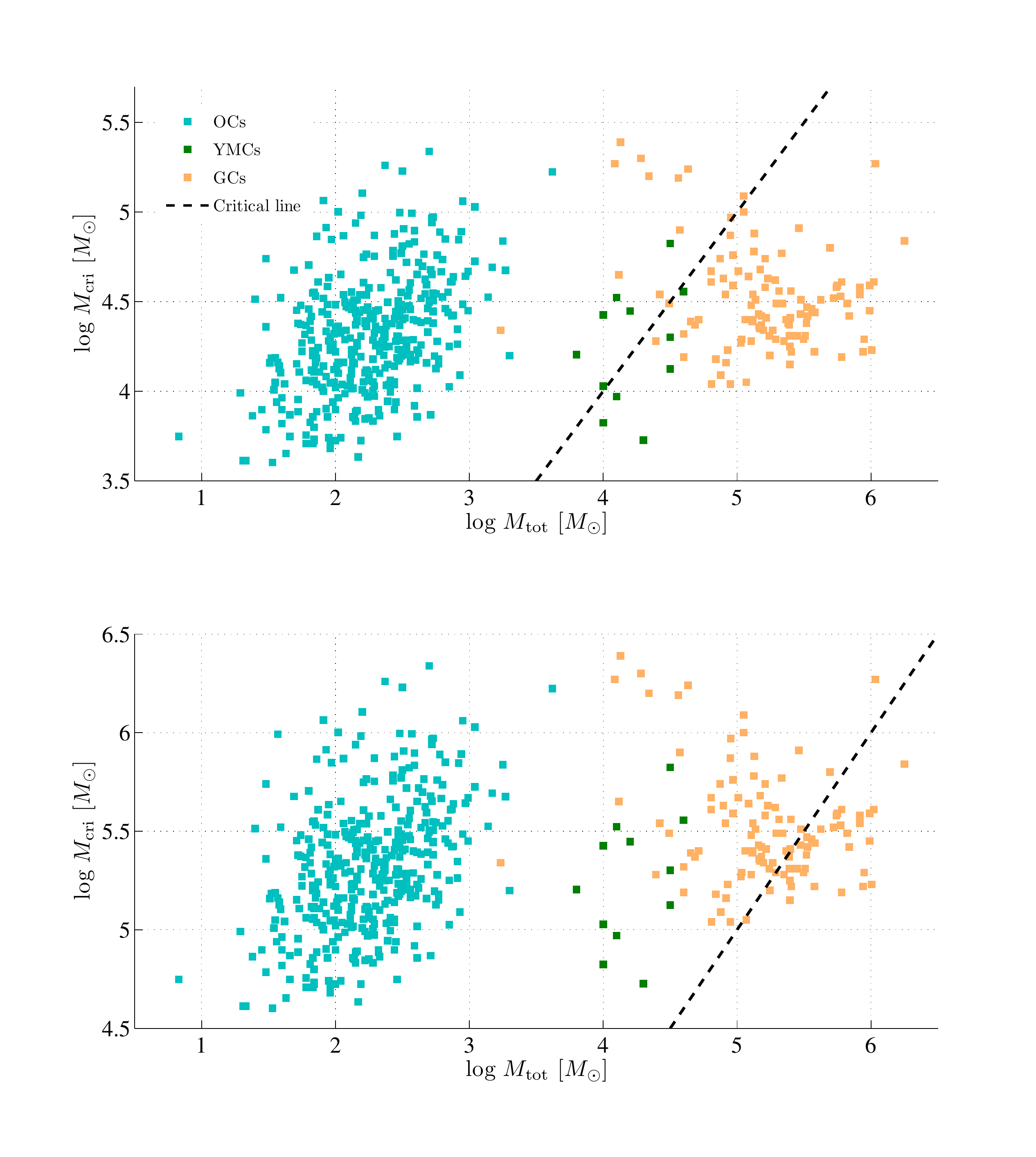}
\caption{Critical mass, $\log M_{\rm cri}$, to capture the initial
  stellar winds of AGB stars for star clusters as a function of their
  actual mass, $\log M_{\rm tot}$. Mint, dark green, and light orange
  rectangles: Galactic OCs, YMCs, and GCs, respectively. OC and YMC
  data are from \citet[][private communication]{Port10a}. GC data are
  derived from
  \cite[][http://www.astro.caltech.edu/\~{}george/glob/glob.data]{Djor94a}. The
  rectangles located above the critical (black dashed) lines are
  clusters which will not be able to retain their initial runaway
  material (based on their current masses). Escape velocities: (top)
  10 km s$^{-1}$; (bottom) 100 km s$^{-1}$.}\label{F1}
\end{figure}

Assuming that the average half-mass radius of young clusters is
$\sim$1 pc, we can evaluate the escape velocity \citep{Geor09a},
\begin{equation}
v_{\rm esc}(t) \approx 0.1\sqrt{\frac{M_{\rm cl}(t)}{r_{\rm
      h}(t)}}\quad {\rm km}\,{\rm s}^{-1}.
\end{equation}
It follows immediately that only for clusters with masses of $M_{\rm
  cl} = 10^4$ $M_{\odot}$ to $10^6 $ $M_{\odot}$, the initial runaway
stellar material will be retained in their gravitational potential
wells.  In Fig. \ref{F1} we present the minimum mass for clusters to
retain their initial runaway material as a function of their current
mass for two different escape velocities: 10 km s$^{-1}$ corresponds
to the gas-expulsion velocity and the slowest AGB ejecta (top panel),
while 100 km s$^{-1}$ corresponds to the lowest escape velocity driven
by Type II supernovae (bottom panel). Figure \ref{F1} shows that the
OCs will fail to capture their initial runaway gas in both cases,
while about half of the observed YMCs in the Milky Way will fail to
capture the slowest AGB ejecta. For an escape velocity of 100 km
s$^{-1}$, all YMCs will lose their initial gas. This also holds for
most GCs. Note that the escape velocity of 100 km s$^{-1}$ is only a
lower limit to the velocities generated by supernova explosions, so
our estimates are conservative.

\

\fbox{%
  \parbox{0.90\textwidth}{%
      This is why star clusters are thought to approximate SSPs:
      during the initial phase ($\leq$ 1 Myr), stellar feedback is
      sufficiently strong to unbind most residual gas from the
      protocluster; O-type stars will subsequently give rise to
      multiple supernova explosions (3 Myr to 10 Myr), cleaning out
      all remaining gas. Afterwards, a surviving cluster will become
      less compact and less massive, thus losing its capacity to
      retain additional gas.
  }%
}  

\

Initial star-formation episodes in clusters cannot have lasted longer
than the timescale of initial gas expulsion, i.e., the
first-generation stars can only have been formed over a short
timescale. In fact, the member stars of extremely young star clusters
(with ages $\le$ 3 Myr) are usually found in regions of their natal
GMCs that are largely devoid of gas, while all YMCs older than several
tens of millions of years are fully exposed.

Numerous studies of young star clusters have confirmed the SSP
scenario: numerical simulations have shown that star formation
proceeds in localized bursts within the cloud \citep{Bate03a,Fari15a},
after which gas expulsion will remove most of the initial molecular
gas. Clusters will also lose many of their member stars at the same
time \citep{Baum07a}. Some nascent, extremely young clusters, commonly
referred to as `embedded clusters' \citep{Bast06a,Davi12a,Lima14a},
are indeed observed. Combined with observations of YMCs with ages of
$\sim$100 Myr, several lines of evidence imply that there will be
little residual gas for star formation in YMCs after the gas-expulsion
period \citep{Bast13a,Bast14a,Cabr15a}.

As discussed in Section \ref{S1}, star clusters will lose stars
through dynamical two-body relaxation. This disruption process is
further accelerated by external tidal shocks. As a consequence, a star
cluster's mass will continue to decrease during its entire evolution
(if no additional merger or accretion events occur), which means that
clusters can retain their initial gas only during the early stages of
their evolution. However, detailed studies have confirmed that OCs and
YMCs are SSPs, with rare exceptions, such as the cluster NGC 6791
\citep{Geis12a}.

The ubiquitous multiple stellar populations in GCs immediately pose a
conundrum: How do these clusters trigger secondary star formation if
most of their initial material would have been lost at an early stage?
In fact, this issue relates to another key open question: Do GCs have
the same origin as young star clusters? In Section \ref{S3} we
introduce some relevant background to acquaint readers with the
details of stellar populations in various types of star clusters.

\section{Not-so-simple Stellar Populations in Star Clusters}\label{S3}
\subsection{Stellar Populations in Young and Intermediate-age Clusters}
Spectroscopic analyses investigating star-to-star chemical variations
provide evidence that OCs are chemically homogeneous
\citep[e.g.,][]{Shen05a,DOra09a,Ting12a}---this seems to also hold for
old OCs like NGC 188, see \cite{Norr85a}---thus confirming that their
stars indeed formed coevally from a common GMC. This conclusion is
also valid for stellar associations \citep[e.g.,][]{DeSi13a} and
star-forming regions \citep[e.g.,][]{Niev11a}, indicating a high
degree of chemical homogeneity in the primordial GMCs.

The most direct method to examine if a group of stars in a star
cluster are coeval is by exploring their CMD, since the CMD of an SSP
can be described very well by a single isochrone. However, because of
Galactic extinction, OC CMDs are usually messy, which renders
explorations of the age distributions of their member stars
difficult. Currently, measuring a single star's age is still
challenging. No single method works well for a broad range of stellar
types or for the full range in ages, but this may change in the next
few years with improvements resulting from the technique of
asteroseismology \citep{Sode10a}.

Another promising approach to studying stellar ages is by measuring
their surface lithium abundances, since the lithium abundance is
suggested to decrease with increasing stellar age. Studies based on
the Li clock surprisingly reveal that the ages of some stars seem to
exceed their host OCs' isochronal ages
\citep{Dunc83a,Pall07a}. However, it is unclear whether these
exceptions can be made to resemble large-scale ongoing star
formation. For instance, \cite{Sacc07a} showed that only three of
their 59 stars analyzed in the young $\sigma$ Orionis cluster show Li
depletion, and among the former only one star has a nuclear age that
exceeds the isochronal age. \cite{Bik14a} analyzed the ages of the
most massive O-type stars in the extremely young, massive cluster W3
Main (with an age of approximately 3--5 Myr). They found that star
formation in W3 Main has lasted for about 2--3 Myr and is still
ongoing.

In addition to the Li clock, which only works for individual stars,
exploring the luminosity distribution of main-sequence turn-off (MSTO)
stars provides an independent approach to test the SSP
scenario. Although Galactic extinction makes studying the intrinsic
magnitude spread of MSTO stars in OCs difficult, some exceptions
exhibit robust statistical results. Based on a comparison of stellar
isochrones and the observed stellar distribution in the CMD,
\cite{Egge98a} found that the age ranges of the Hyades and Praesepe
OCs seem to span several hundred million years, as implied by the
apparent luminosity spread of MSTO stars. A similar luminosity
dispersion has also been found in the Orion Nebula Cluster, for which
an internal age spread on the order of 10 Myr was inferred
\citep{Pall00a}. More recent studies have found that NGC 1856, a 300
Myr-old Large Magellanic Cloud (LMC) cluster, displays an apparently
extended MSTO (eMSTO) region \citep{Milo15a}, which may indicate
prolonged star formation, lasting $\sim$150 Myr. A recent study found
that the 150 Myr-old YMC NGC 1850 also seems to harbor an eMSTO region
\citep{Bast16a}. Because both NGC 1850 and NGC 1856 reside in the LMC,
they are not affected by significant extinction due to dust in the
Galactic disk. The morphologies of the eMSTO regions in these YMCs may
be a `smoking gun,' suggesting that stars in star clusters may form
continuously rather than in a single starburst event.

Such eMSTOs seem to be ordinary features of intermediate-age (1--2
Gyr-old) star clusters in the LMC and the Small Magellanic Cloud
(SMC). For instance, \cite{Mack07a} found that the LMC cluster NGC
1846 shows a split in its MSTO region, which can be well described by
two isochrones with ages of 1.5 Gyr and 1.8 Gyr. Similar results have
been found for the LMC clusters NGC 1783 and NGC 1806, with
indications that their age dispersion may be at least 300 Myr
\citep{Mack08a}. \cite{Milo09a} studied 16 intermediate-age LMC
clusters by exploring their MSTO regions, concluding that 70\% have
MSTOs that are inconsistent with the expectations from
SSPs. \cite{Gira13a} studied the SMC clusters NGC 411 and NGC
419. They found that their eMSTO regions are consistent with age
spreads of $\sim$700 Myr, which may be the largest apparent age
spreads known among all intermediate-age star clusters \citep[but
  see][]{Wu16}. Similar results have also been found by
\cite{Rube10a,Rube11a}. \cite{Gira09a} even claimed the presence of
dual red clumps (RCs) in NGC 419. \cite{Li14a} analyzed the MSTO
regions in the LMC clusters NGC 1831 and NGC 1868. They found that
their implied internal age spreads are 280 Myr and 320 Myr,
respectively, although the compact RC of NGC 1831 is inconsistent with
such a large age spread.
 
The discovery of eMSTO regions in intermediate-age star clusters
strongly challenges our understanding of star cluster formation and
evolution. Various scenarios have been proposed to explain the
observations. \cite{Goud11a,Goud14a} found a correlation between the
widths of the MSTO regions and the early escape velocities in
intermediate-age star clusters. They claimed a velocity threshold of
12--15 km s$^{-1}$ for all clusters featuring eMSTOs, which is
consistent with the escape velocity of gas expulsion and slow AGB
stellar winds. This implies that the gas used for secondary star
formation may have originated from intermediate-mass AGB stars or
massive binary stars. However, since most of these star clusters are
more massive than 10$^4$ $M_{\odot}$, their initial stellar
populations should have contained many massive OB stars. It is then
concerning that these OB stars did not expel the interstellar medium
at very early stages through stellar feedback and supernova
explosions. This scenario was challenged by \cite{Cabr16a}, who
carefully studied the stellar populations of NGC 7252-W3, a possible
YMC candidate with an age of $\sim$570 Myr and an escape velocity of
193 km s$^{-1}$. However, no evidence of an extended star-formation
history (eSFH) was found. \cite{Bast14a} searched for residual gas in
13 LMC and SMC YMCs with ages $\leq$300 Myr. Their conclusion was,
once again, negative: No residual gas was detected. In summary, most
clusters are unable to accrete large gas reservoirs to support ongoing
star formation. However, this does not mean that they will not be able
to capture external gas after the initial gas-expulsion phase.
\cite{Li16a} found the clear presence of younger stellar populations
in three intermediate-age star clusters, NGC 1783, NGC 1806, and NGC
411, which seem to have an external origin. These young stellar
populations are tightly associated with SSP isochrones, which cannot
be explained by contamination due to field stars spanning a range of
ages and metallicities. In addition, the radial profiles of these
stellar populations clearly peak in the clusters' central regions (for
a more detailed rebuttal, see Section \ref{S4}). They speculated that
these star clusters may have accreted external gas to form new stars
at a later stage, although the newly born stars only represent a minor
fraction in mass.

\begin{figure}[htbp]
\centering
\includegraphics[width=4.0in]{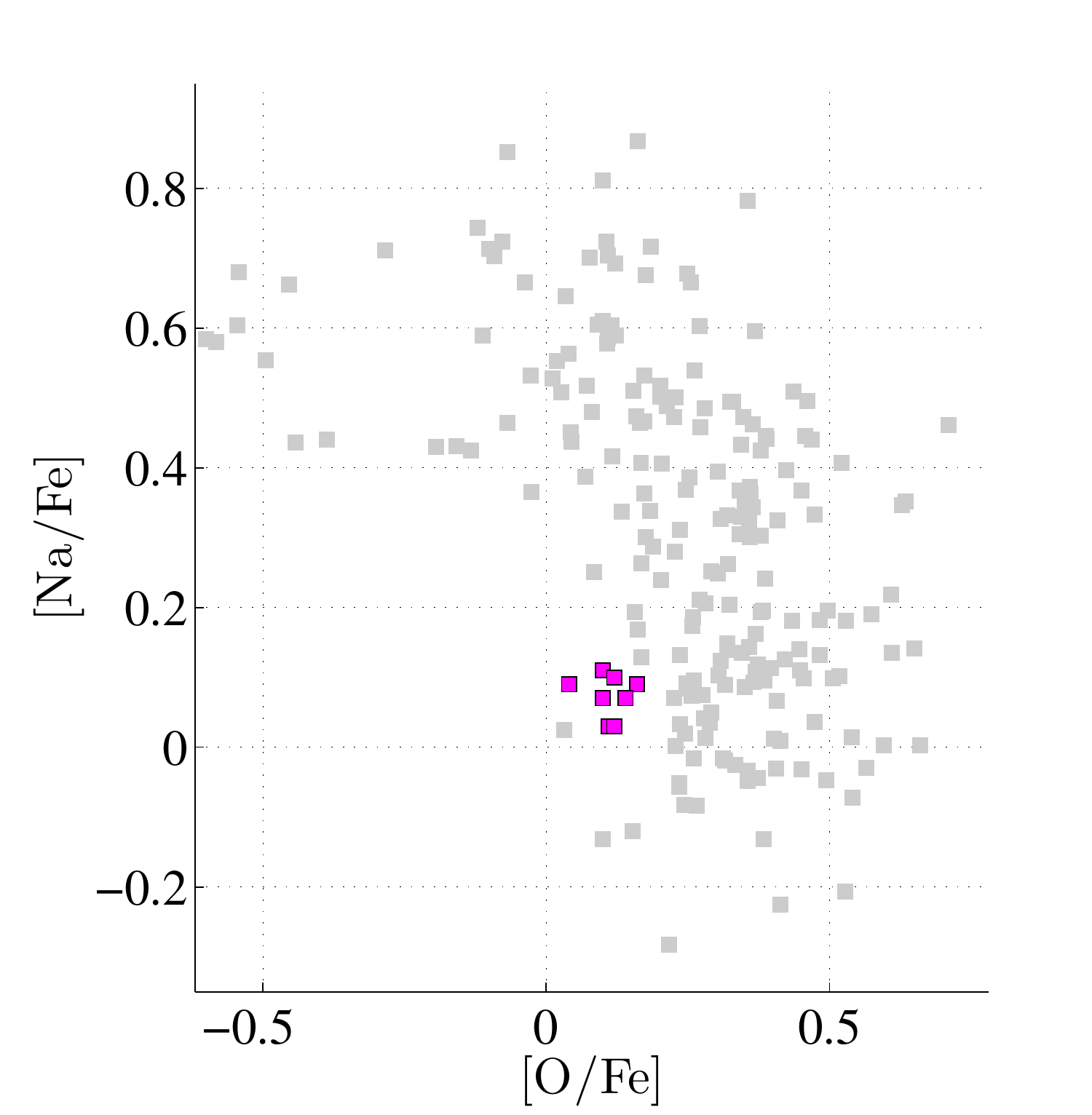}
\caption{Na--O diagram of stars in NGC 1806 (pink) and Galactic GCs
  (gray). The NGC 1806 data were obtained from \cite{Mucc14a}; GC data
  are from \cite{Carr09a}.}\label{F2}
\end{figure}

As regards the eSFH problem, numerous debates now focus on whether
other features of intermediate-age star clusters are also consistent
with an eSFH. \cite{Li14b} found that in the 1.7 Gyr-old cluster NGC
1651, despite exhibiting an eMSTO consistent with a 450 Myr age
spread, the SGB morphology can only be reconciled with an age spread
of $\leq$160 Myr. The same conclusion was reached by \citet{Bast15a}
for NGC 1806 and NGC 1846. However, their results were criticized by
\cite{Goud15a}, who performed numerical simulations based on these
clusters' physical parameters to show that the morphologies of their
SGBs may still be consistent with significant age spreads. To explain
the apparent inconsistencies between the widths of the SGBs and the
stellar evolutionary models, they suggested stellar convective
overshooting as a viable alternative solution. However, \cite{Li15a}
found a more conspicuously narrow SGB in NGC 411, which can only be
explained by an SSP. This is in obvious contrast with the extent of
its eMSTO region. They concluded that NGC 411 is more likely composed
of a coeval stellar population rather than of multiple-aged stellar
populations. \cite{Gira13a} claimed that NGC 411 appears to have
experienced an eSFH lasting $\sim$700 Myr, which immediately
introduces an apparent discrepancy with respect to the observed
SGB. On the other hand, \cite{Mucc14a} studied eight giant stars in
the cluster NGC 1806 and did not find any evidence of a possible Na--O
anticorrelation, common to almost all Galactic GCs. The discovery of
\cite{Mucc14a} indeed stands in marked contrast to the chemical
variations present in GCs. In Fig. \ref{F2} we present the Na--O
diagram pertaining to the NGC 1806 stars (pink), compared with that of
GCs (gray).

\subsection{Stellar Populations in Globular Clusters}
Old GCs appear to represent a separate population of star clusters
compared with OCs, YMCs, and also intermediate-age star
clusters. Multiple stellar populations are thought to be a common
feature of GCs in the Milky Way \citep{Grat12a}. The first evidence of
multiple stellar populations in GCs may date back to the 1980s:
\cite{Hess80a} found a dispersion in the CN-band strengths among seven
MS stars in the GC 47 Tucanae (47 Tuc), indicating that a process
leading to N enhancement must have occurred in some of those
stars. Since MS stars in 47 Tuc lack the capacity of mixing N to the
stellar surface, such a CN variation must have a primordial
origin. Similar results were also confirmed in other GCs, e.g., in NGC
362 \citep{Smit83a}, NGC 6171 \citep{Smit88a}, M71 \citep[NGC
  6838;][]{Penn92a}, M3 \citep[NGC 5272;][]{Smit02a}, and M13
\citep[NGC 6205;][]{Smit06a}). N-enhanced stellar populations exhibit
a strong absorption line at $\sim$3400 {\AA}: see \citep[][their
  Fig. 11]{Milo12a} and \citep[][their Fig. 1]{Piot15a}. Variations of
the N abundance in stellar populations will affect their CMD
morphologies in ultraviolet filters, producing a split in the
RGB. Such a feature has been detected in the CMDs of a dozen GCs
\citep{Piot15a}.

Other evidence of multiple stellar populations in GCs comes from the
variations in light elements (Na, O, Mg, Al), including the Na--O
anticorrelation: see Fig. \ref{F2}. For more details on light-element
chemistry in GCs, we recommend the review by \cite{Grat12a}. The Na--O
anticorrelation in GCs is exhibited by MSTO, SGB, and RGB stars
\citep{Carr04b,Grat04a}. It has been suggested to result from the CNO
cycle ($\sim 20 \times 10^{6}$ K) and the Na proton-capture process
($\sim 35 \times 10^6$ K). The central temperatures of the MSTO stars
in GCs ($\sim$ 0.85 $M_{\odot}$) do not reach the threshold
temperature for initiation of the proton-capture process required for
the production of Na \citep{Decr07a}. Therefore, the Na--O
anticorrelation is unlikely the result of stellar evolution
\citep[e.g., processes affected by the first dredge-up or meridional
  mixing;][]{Swei79a}, but it must instead have a primordial origin
intrinsic to the nature of multiple stellar populations. A similar
Mg--Al anticorrelation has also been found in GCs
\citep[e.g.,][]{Grat01a,Yong03a}; for more details, see
\cite{Grat04a,Grat12a}. It seems that all Galactic GCs harbor Na--O
anticorrelations, while light-element variations are also present in
some LMC GCs \citep{John06a}.

The iron abundance in a star cluster can only be enhanced through the
ejecta of massive, core-collapse supernovae
\citep{Cayr86a,Parm99a}. If a cluster exhibits an iron dispersion,
this indicates that its SFH has lived through supernova
explosions. Such clusters have been observed: \cite{Mari09a} explored
17 giant stars in the GC M22 and identified two groups with different
iron and s-process abundances. A similar result, for M2, was obtained
by \cite{Yong14a}, who determined a metallicity dispersion exhibiting
three peaks, at [Fe/H] $\sim -1.7,-1.5$, and $-1.0$ dex. Such a
metallicity dispersion has also been found in the GC NGC 1851
\citep{Carr10a}. The examples of M2, M22, and NGC 1851 strongly
suggest that at least some GCs have experienced much more complex SFHs
than provided by individual starburst episodes.

More impressive evidence comes from precise photometry: the multiple
stellar groups that show different properties in the CMD have also
been found to possess different chemical compositions. For example,
NGC 1851 was found to harbor dual SGBs; its two branches were also
found to have different CNO contents \citep{Cass08a}. \cite{DAnt05a}
studied the color distribution of MS stars in the GC NGC 2808 based on
{\sl Hubble Space Telescope} ({\sl HST})/Wide Field Planetary Camera-2
(WFPC2) observations. They found that the width of its MS is
inconsistent with an SSP, which requires a helium abundance ranging
from $Y=0.26$ to $Y=0.40$. This has been confirmed by \cite{Piot07a},
who obtained accurate photometry of NGC 2808 based on observations
with the {\sl HST}/Advanced Camera for Surveys (ACS). They found that
the broadened MS is actually composed of three distinct subsequences,
indicating three stellar populations with helium-abundance peaks at
$Y=0.248$, $Y=0.30$, and $Y=0.37$. Spectroscopic evidence was provided
by \cite{Mari14a}, who analyzed 96 HB stars in NGC 2808 and found that
their Na content depends on their position in the CMD, with blue HB
stars having higher Na abundances than their red counterparts. The
multiple features in GCs are more apparent in ultraviolet observations
\citep{Piot15a}. A recent study of NGC 7089 (M2) even revealed seven
stellar populations \citep{Milo15b}.

Photometric analyses also enable one to investigate the kinematics and
dynamics of the different stellar populations. \cite{Milo12a} found
dualities in the MS, SGB, and RGB of 47 Tuc, which can be explained by
a CN-weak, O-rich, and Na-poor first stellar generation with a normal
helium abundance ($Y\sim 0.25$) and a secondary stellar generation
with CN-strong, O-poor, Na-rich elemental abundances as well as an
enhanced helium abundance ($Y\sim 0.265$). The radial behaviors of
both stellar generations are different: the second stellar generation
is more concentrated than the first generation. \cite{Li14c} placed
their results on a firmer footing through near-infrared
observations. They determined more obvious radial stellar population
gradients, concluding that the central region of 47 Tuc is almost
entirely dominated by the secondary stellar generation, while the
cluster's periphery is fully composed of first-generation stars. They
also found that the secondary stellar generation has enhanced helium
abundance and metallicity. Similar results were also found for NGC 362
\citep{Carr13a} and several other GCs \citep{Lard11a}.

The observational status of stellar populations in star clusters can
be summarized as follows:
\begin{itemize}
\item[(a)] Despite small numbers of stellar members, extremely young,
  massive clusters still embedded in their GMCs (usually with ages
  $\leq$3 Myr) are characterized by proto-stellar populations that
  formed in a single episode with a very small age spread.
\item[(b)] Intermediate-age LMC and SMC star clusters (but also
  including NGC 1850 and NGC 1856, which have ages of only 150 Myr and
  300 Myr, respectively) exhibit eMSTO regions. If their extents are
  interpreted as fully caused by a range in stellar ages, they are in
  apparent conflict with the SSP scenario. However, no evidence of
  chemical dispersions has been found in these clusters
  \citep{Mucc14a}. The tight SGBs in some of these clusters also
  contradicts a significant age spread
  \citep{Li14b,Li15a,Bast15a}. Whether these clusters may be
  characterized by initially extended SFHs is still being debated.
\item[(c)] {\sl The presence of multiple stellar populations in GCs is
  undisputed, making it the greatest open challenge to the SSP
  scenario for star clusters.} A combination of spectral and
  photometric evidence supports the presence of multiple stellar
  populations in GCs, which also differ in both their kinematics and
  dynamics. This may indicate primordial chemical
  inhomogeneities. However, as we will show in Section \ref{S4},
  stellar evolution is probably still a straightforward solution to
  explain the observations of some of these clusters.
\end{itemize} 

\section{Proposed Scenarios}\label{S4}
\subsection{Controversy about Multiple Stellar Populations in Young and Intermediate-age Clusters}
An intuitive explanation of the eMSTO regions observed in
intermediate-age star clusters is that their member stars may have
formed during a time span of several hundred million years. Since the
morphology of the MSTO is very sensitive to the age distribution of
its stellar population, many authors have suggested that eMSTO regions
in intermediate-age clusters might imply eSFHs of {\sl at least} 300
Myr. \cite{Goud14a} proposed a model for massive clusters
  characterized by a prolonged round of star formation: the central
  escape velocities of their eMSTO model star clusters were
  sufficiently high to accrete a significant amount of pristine gas
  from their surroundings or to retain the gas from the ejecta of
  intermediate-mass AGB stars and massive binaries. Since the remnant
  gas does not fully escape from the infant clusters, most of this gas
  reservoir would be accumulated in the central regions of the
  clusters, subsequently forming secondary populations of stars. This
  process would last several million years until all gas is exhausted
  by star formation or multiple supernova explosions. However, the
lack of residual gas in YMCs \citep{Bast13a} all but invalidates this
scenario. A second proposed origin of such an age spread is the merger
between two clusters with different ages \citep{Mack07a} or between a
cluster and a star-forming GMC \citep{Bekk09a}. If this
  scenario turns out to work, it would imply that most LMC clusters
  actually belonged to binary systems in the past, since 70\% of
  intermediate-age star clusters in the LMC show such eMSTOs
  \citep{Milo09a}. However, presently binary clusters only account for
  about 10\% of the current cluster population of the Magellanic Clouds \citep{Dieb00a}.

An alternative scenario to explain eMSTOs draws on a range in stellar
rotation rates. Stellar rotation can alter the morphology of MSTO
regions in two ways: (i) the centrifugal force resulting from rotation
reduces the stellar self-gravity, thus decreasing the surface
effective temperature, which in turn renders its observed color redder
than that of the non-rotating counterparts. The reduction in stellar
self-gravity also decreases the stellar core pressure, hence reducing
the central nuclear reaction rate \citep{Faul68a}, which causes a
dimming of fast rotators compared with slowly or non-rotating
stars. The observed stellar luminosities and colors also depend on the
inclination angles of individual stars: a rapidly rotating star will
look redder around its equatorial region than near its poles. This
effect is known as `gravity darkening' \citep[see][their
  Fig. 1]{Geor14a}. (ii) Rotation enlarges the stellar convective
cores, transporting hydrogen from the outer layers into the central
region, in turn replenishing the central fuel for hydrogen burning,
which may thus prolong the stellar MS lifetime. This is called
`rotational mixing.' It has been suggested that stellar rotation of
O-, early-B-, and F-type stars may lead to lifetime increases by about
25\% \citep{Maed00a,Gira11a}.

Less massive stars ($\leq 1.2 M_{\odot}$) are not expected to become
fast rotators because of magnetic braking \citep{Scha62a,Mest87a},
i.e., the stellar magnetic field exerts a torque on the ejected matter
during stellar evolution, resulting in a steady transfer of angular
momentum away from the star. For example, stars earlier than F0-type
can easily reach an average rotational velocity of 100--200 km
s$^{-1}$, while the typical rotational velocity of G0-type stars is
only 12 km s$^{-1}$ \citep{McNa65a}. A more recent study of a large
number of B--F-type solar-neighborhood stars was carried out by
\cite{Roye07a}, who confirmed that most of these stars are fast
rotators.

\begin{figure}[htbp]
\centering
\includegraphics[width=5.0in]{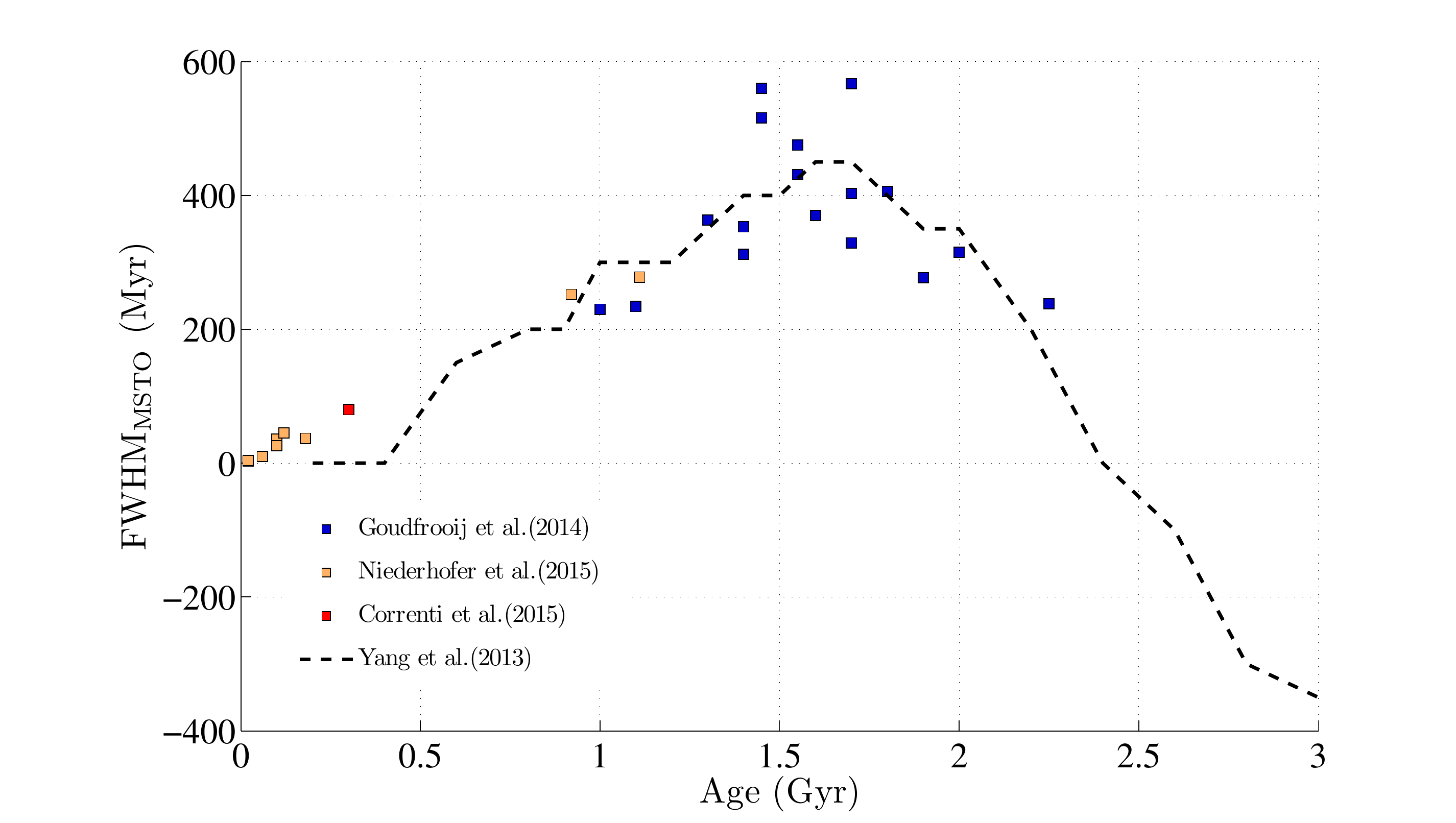}
\caption{Widths of implied cluster age spreads as a function of
  isochronal age. Black dashed line: Predicted FWHM of cluster age
  spreads that would be derived from their eMSTO regions as a function
  of cluster age. A positive (negative) FWHM means that the
  populations of fast rotators are redder (bluer) than the non- or
  slowly rotating populations. Blue, red, and orange rectangles: star
  cluster data from \cite{Goud14a}, \cite{Nied15b}, and
  \cite{Corr15a}, respectively.}\label{F3}
\end{figure}

\cite{Bast09a} suggested that rapid stellar rotation of F-type stars
may lead to the misconception that intermediate-age star clusters
harbor significant age spreads. In their scenario, the dominant effect
that is responsible for the extent of the (e)MSTO region is gravity
darkening. However, \cite{Gira11a} countered that their model is
unrealistic, because it ignores rotational mixing. The latter authors
showed that rotational mixing will mitigate the broadening caused by
gravity darkening. As a result, the combination of these two effects
may still produce a narrow MSTO, and therefore, they argued, stellar
rotation cannot be the (only) cause of the eMSTO. However,
\cite{Yang13a} pointed out that the conclusion of \cite{Gira11a} is
tenable only in the presence of a high convective mixing
efficiency. \cite{Yang13a} studied the collective effects of gravity
darkening and rotational mixing with a moderate mixing
efficiency. They successfully reproduced the eMSTO regions for 1--2
Gyr-old clusters. A corollary of their results is that the MSTO area
in a cluster depends on its actual age \citep[see
  also][]{Bran15a}. \cite{Nied15a} found a correlation between the
full width at half maximum (FWHM) of the postulated age spreads in
clusters and their actual ages (based on isochrone fitting to the blue
boundary of their eMSTO regions) for 12 intermediate-age LMC star
clusters, which is consistent with the predictions of the stellar
rotation scenario. In Fig. \ref{F3} we present the predicted
correlation between the FWHM of the clusters' age spreads and their
typical ages for coeval stellar populations \citep[reproduced from
  Fig. 8f of][]{Yang13a}; observational data are also included
\citep{Goud14a,Corr15a,Nied15b}.

Figure \ref{F3} shows that the MSTO regions of clusters with ages of
up to 2.5 Gyr are more extended than expected for SSPs. The maximum
age spread derived from the eMSTO region occurs for ages ranging from
$\sim$1.5 Gyr to 2.0 Gyr. In \cite{Yang13a}'s model, a positive FWHM
means that gravity darkening is the dominant effect, making the fast
rotators redder than their non- or slowly rotating counterparts. On
the other hand, a negative FWHM implies that rotational mixing
dominates, resulting in fast rotators to appear bluer than the non- or
slow rotators. \cite{Bran15a} also found that the area of a cluster's
MSTO region depends on its age if its MSTO stars have different
rotation rates. However, their models imply that rotational mixing is
likely the dominant effect responsible for the extent of the eMSTO
region at any age, i.e., the fast rotators are bluer than the non- or
slow rotators, predominantly owing to rotational mixing. The maximum
age spread derived from their MSTO regions peaks at $t \in$ [1.0,1.5]
Gyr.

If the apparently significant age spreads of several hundred million
years are indeed valid for YMCs ($\leq$ 300 Myr), one should expect
pre-MS stars to appear in YMC CMDs, i.e., the age distributions of
their member stars would scatter to zero age. However, as shown in
Fig. \ref{F3}, the internal age spreads of YMCs derived from their
eMSTO regions only span small fractions of their ages. Once again,
this supports the rapid stellar rotation scenario, since it will only
partially broaden the MSTO region.

The relative importance of gravity darkening and rotational mixing in
star clusters is as yet unclear owing to a lack of direct
observational evidence. The main difference between these processes
resides in the loci of the rapidly rotating population. Gravity
darkening will produce an eMSTO where most fast rotators reside toward
the red side of the MSTO, while rotational mixing will result in fast
rotators being distributed on the blue side of the MSTO
region. However, measuring stellar rotation in dense star clusters at
the distance of the LMC is difficult with current facilities. Since
both stellar rotation and age spreads will produce eMSTOs in star
clusters, important differences which may allow us to distinguish
between these processes may appear on the SGB. Because gravity
darkening does not produce a mass spread among MSTO stars, once the
MSTO stars have evolved off the MS, they will slow down rapidly as a
result of the conservation of angular momentum. Subsequently, the
coeval MSTO population characterized by different stellar rotation
rates will converge into a narrow SGB \citep{Wu16}. However,
rotational mixing will produce a mass spread among evolved stars,
which will still broaden both the MSTO and SGB regions. In Table
1 we summarize the features of the MSTO region and the SGB for
four cases: (1) an SSP without stellar rotation; (2) a significant age
spread (eSFH); (3) an SSP with fast rotators affected by gravity
darkening; and (4) an SSP with fast rotators affected by rotational
mixing.

\begin{table}[tbp]\label{T1}
\caption{MSTO and SGB features corresponding to (1) an SSP, (2) an age
  spread (eSFH), (3) gravity darkening, and (4) rotational mixing.}
\centering
\begin{tabular}{lcc}
\hline
Scenario          & MSTO Region                      & SGB Region \\ \hline  
SSP               & Narrow                           & Narrow \\         
eSFH              & Extended in color and luminosity & Extended in luminosity \\
Gravity darkening & Extended in color and luminosity & Narrow \\        
Rotational mixing & Extended in color and luminosity & Extended in luminosity\\ \hline
\end{tabular}
\end{table}

\begin{figure}[htbp]
\centering
\includegraphics[width=6.0in]{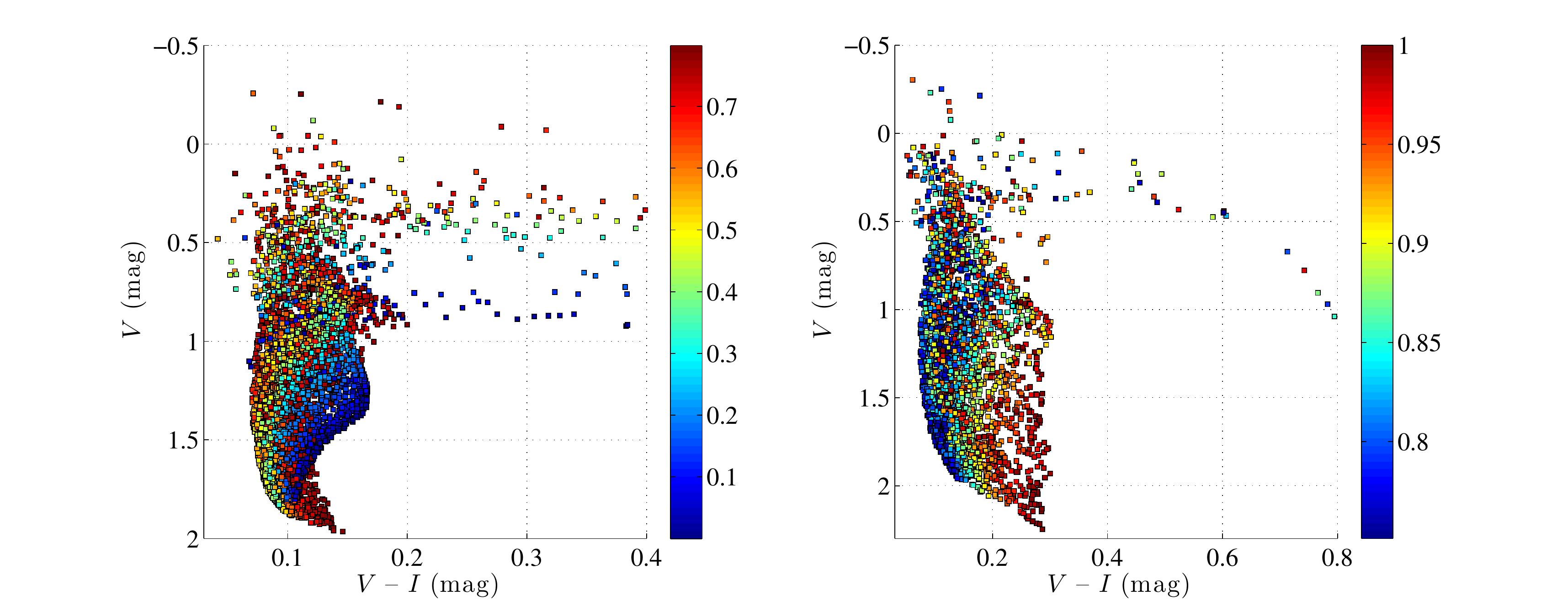}
\caption{Synthetic ($V, V - I$) CMD for a coeval stellar population
  (based on the Geneva models) evolved to an age of 1 Gyr and with a
  metallicity of $Z = 0.006$, for stars with initial rotation rates of
  (left) $\Omega_{\rm ini}/\Omega_{\rm crit}$ $\leq$ 0.7 and (right)
  $\Omega_{\rm ini}/\Omega_{\rm crit}$ $\ge$ 0.7. The color bars
  represent the initial stellar rotation rates.}\label{F4}
\end{figure}

The Geneva stellar evolutionary code includes the effects of initial
stellar
rotation,\footnote{\url{http://obswww.unige.ch/Recherche/evol/-Database}}\citep{Ekst12a,Geor13a,Geor13b,Yuso13a}
but it currently only covers a limited stellar mass range and a small
number of fixed initial rotation rates. (Observationally, we only have
access to current stellar rotation rates).\footnote{The Geneva models
  are only available for nine different initial rotation rates,
  $\Omega_{\rm ini}/\Omega_{\rm crit}$ = 0.0, 0.1, 0.3, 0.5, 0.6, 0.7,
  0.8, 0.9, and 0.95, where $\Omega_{\rm ini}$ is the initial stellar
  angular rotation rate and $\Omega_{\rm crit}$ is the critical,
  `break-up' value. The stellar masses included range from $1.7
  M_{\odot}$ to $15 M_{\odot}$. In addition, for the mass range
  between $0.8 M_{\odot}$ and $120 M_{\odot}$, the model suite is
  defined for two initial rotation rates, $\Omega_{\rm
    ini}/\Omega_{\rm crit} = 0.0$ and 0.568.} In Fig. \ref{F4} we
present synthetic CMDs of SSPs with $\Omega_{\rm ini}/\Omega_{\rm
  crit}$ $\leq 0.7$ (left) and $\Omega_{\rm ini}/\Omega_{\rm crit}$
$\ge 0.7$ (right) evolved to an age of 1 Gyr and for a metallicity of
$Z = 0.006$, i.e., the value closest to that typical of most
intermediate-age LMC and SMC star clusters. Application of the Geneva
code shows that for initial stellar rotation rates of $\Omega_{\rm
  ini}/\Omega_{\rm crit}$ $\leq$ 0.7, rotational mixing is the
dominant effect responsible for broadening the MSTO region: the
relatively rapid rotators are bluer than their slowly rotating
counterparts. In contrast, for stars with $\Omega_{\rm
  ini}/\Omega_{\rm crit}$ $\ge$ 0.7, the fast rotators are redder than
the slow rotators (which also have initial rotation rates $\Omega_{\rm
  ini}/\Omega_{\rm crit} \ge$ 0.7), indicating a dominant importance
of gravity darkening among the rotating stellar population. A
remarkable difference appears on their SGBs, since rotational mixing
produces a mass spread for evolved stars, which thus causes a
broadened SGB. However, as shown in the right-hand panel of
Fig. \ref{F4}, gravity darkening will not broaden the SGB.

\begin{figure}[htbp]
\centering
\includegraphics[width=6.0in]{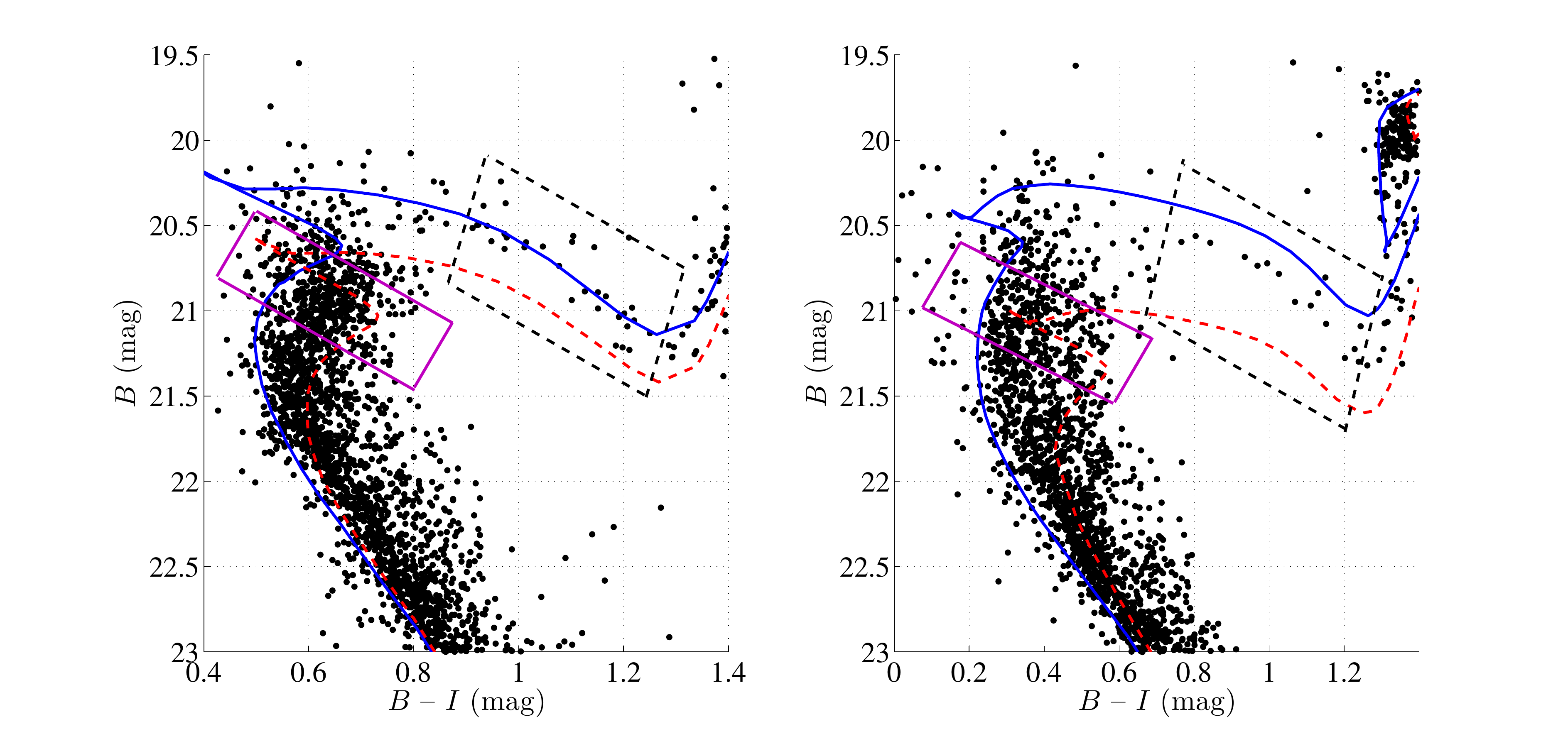}
\caption{CMDs of the intermediate-age star clusters NGC 1651 (left)
  and NGC 411 (right). Both harbor eMSTO regions (purple outlines) and
  well-populated SGBs (black dashed outlines). The best-fitting
  isochrones (blue, red solid lines: young, old isochrones) are also
  shown, determined based on the extents of their eMSTO
  regions.}\label{F5}
\end{figure}

A breakthrough was achieved by \cite{Li14b} and \cite{Bast15a}. Both
independent groups found that the SGB morphologies in three star
clusters, NGC 1651, NGC 1805, and NGC 1846, were inconsistent with
significant age spreads. However, their conclusions were subsequently
challenged by \cite{Goud15a}, who argued that if one were to consider
the effects of stellar convective overshooting, the morphologies of
the clusters' SGBs may still be consistent with the presence of
significant age spreads. The debate thus turned to whether
overshooting is important for F-type SGB stars. \cite{Colu12a} used a
sample of young and intermediate-age LMC clusters to evaluate the
importance of overshooting. They found that the use of isochrones {\bf
  without} convective overshooting results in Fe abundances that most
closely match the loci of stars in clusters with ages between 0.05 Gyr
and 3 Gyr. Another striking contrast between the widths of the eMSTO
region and the SGB is exhibited by the SMC cluster NGC 411, which
hosts an eMSTO region that implies an apparent age spread of $\sim$800
Myr, but it harbors a extremely tight SGB, which can only be described
by a single-aged isochrone \citep{Li15a}. In Fig. \ref{F5} we present
the CMDs of (left) the LMC cluster NGC 1651 and (right) the SMC
cluster NGC 411. The best-fitting isochrones that cover their eMSTO
regions (indicated by the purple outlines) are also shown; the SGB
regions are indicated by regions traced by black dashed lines. One can
immediately see that, compared with their eMSTO regions, the clusters'
SGBs do not show the expected corresponding broadening or
bifurcation. This combined analysis of cluster eMSTO regions and tight
SGBs indeed disfavors the age-spread scenario. Instead, only an
extremely rapidly rotating population will produce an eMSTO combined
with a tight SGB (see the comparison in Fig. \ref{F4}), which may
indicate that stellar rotation in star clusters occurs at much higher
velocities than previously expected \citep{Wu16}.

The discovery of eMSTOs in YMCs further confirms this. The eMSTO of
NGC 1856, a 150 Myr-old YMC, suggests an $\sim$80 Myr age spread
\citep{Corr15a}. \cite{Corr15a} concluded that if the observed eMSTO
is caused by stellar rotation instead, then rotational mixing will
reproduce the observations if the stellar population is composed of a
combination of two-thirds rapidly rotating stars and one-third
slowly/non-rotating stars. The same conclusion was reached by
\cite{Milo16a}, who discovered a split main sequence in the YMC NGC
1755. This feature favors the variable stellar rotation scenario
rather than an age spread. They considered whether all stars in NGC
1755 were born rapidly rotating. All these studies were based on the
premise that rotational mixing is the dominant effect in determining
the extent of an eMSTO. If, on the other hand, stellar rotation were
responsible for the eMSTOs of YMCs, rotational mixing should be of
significant importance. Thus, to further constrain the internal age
distributions and the relative importance of rotational effects,
future investigations of the SGBs or RCs of intermediate-age star
clusters are urgently required.

In summary, the apparent eMSTO regions of intermediate-age star
clusters and YMCs in the LMC and SMC can be explained by scenarios
postulating either an age spread or rapid stellar rotation. The field
still has a long way to go before reaching a unanimous conclusion.

\subsection{Possible Origins for Multiple Stellar Populations in Globular Clusters}
Unlike for intermediate-age star clusters, the presence of multiple
stellar populations in old GCs is well-established. Scenarios that
have been proposed to explain the formation of multiple stellar
populations in old GCs are diverse, but most can be grouped into two
categories. The first type of hypothesis suggests a primordial origin,
in the sense that GCs may have been born with chemical
inhomogeneities, i.e., GCs were formed with a primordial chemical
dispersion. The second type of scenario favors the SSP origin, and the
multiple stellar populations in GCs are then the result of stellar
evolution \citep[see also][]{Kraf94a}.

Most primordial scenarios draw on self-pollution of intra-cluster gas,
occurring during the early stages of cluster evolution. Various
polluters have been proposed, including those originating from the
ejecta of rapidly rotating massive stars \citep{Decr07a}, massive
binaries \citep{deMink09a}, or evolved post-giant-branch stars
\citep{Vent09a}. \cite{Valc11a} proposed a scenario that divides GCs
into three groups, where the most massive GCs, with initial masses
$\geq 10^9 M_{\odot}$ (e.g., M22), can retain the ejecta of all types
of massive stars, including those of their core-collapse supernovae.
Intermediate-mass GCs, with masses of several $\times 10^8 M_{\odot}$
(like NGC 2808) would only be able to retain a fraction of the fast
winds from massive stars. The least massive GCs would not be able to
retain any gas ejected by massive stars (or their products), but they
could form new stars from the slow winds of intermediate-mass stars of
the first stellar generation. Self-enrichment scenarios usually
postulate multiple episodes of star formation. If confirmed, this
means that we may have underestimated the capacity of GCs to collect
and retain gas. Recently, \cite{DAnto16a} proposed a similar
scenario. They claimed that a temporal sequence of AGB generations may
be responsible for the observed multiple stellar populations in GCs:
the pollution process occurs after the Type II supernovae epoch,
lasting until the third dredge-up associated with the AGB
population. They ascribe the cluster-to-cluster abundance variations
to differences in many processes and gas sources which were involved
in the formation of the secondary generation. Based on this scenario,
after the Type II supernovae epoch ($\sim$20 Myr), GCs should still
have been 100 times more massive then their current masses; otherwise
the self-pollution scenario would be hard pressed to explain the high
fraction of secondary stellar generations observed.

\cite{Bast15b} carefully studied all current self-enrichment scenarios
by comparing them to observations. They found that an intrinsic
problem among all these scenarios is that they are unable to produce
consistent abundance trends among He, Na, and O. To form a secondary
stellar generation, clusters should be able to retain or accrete
material from their surroundings \citep{Conr11a}, but current
investigations have shown that clusters are almost gas-free at a very
early stage (2--3 Myr), independent of their mass
\citep{Bast14b,Holl15a}. While these results suggest that clusters may
not be able to retain their gas during the initial gas-expulsion
phase, this scenario does not prevent a cluster from accreting
external gas onto its core potential while moving through a background
medium \citep{Naim11a}.

\begin{figure}[htbp]
\centering
\includegraphics[width=5.0in]{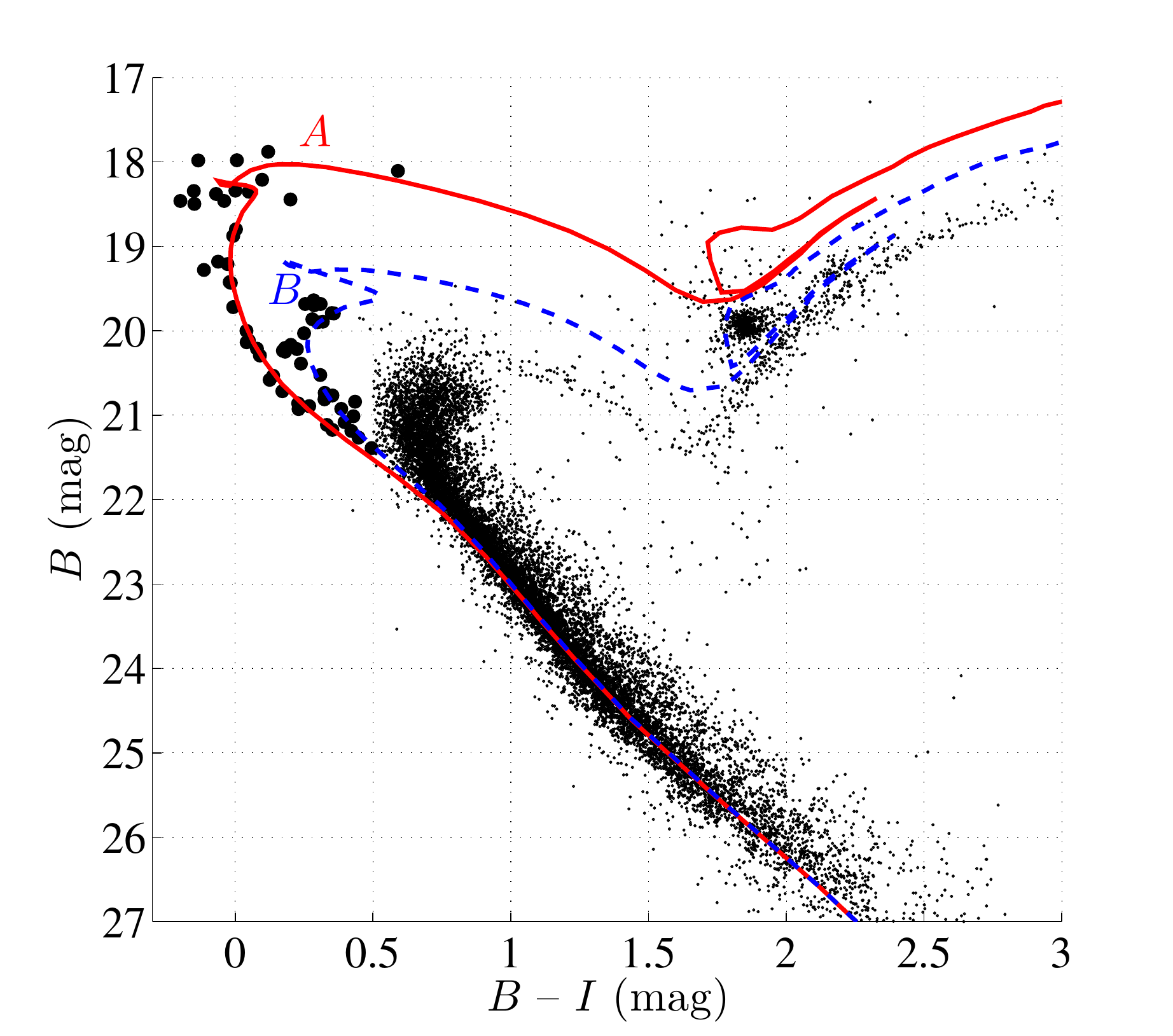}
\caption{CMD of the LMC cluster NGC 1783 (1.4 Gyr old). NGC 1783 hosts
  two apparently younger stellar populations, which are tightly
  associated with two isochrones (shown as the blue dashed and the red
  solid lines), indicated as populations A and B. Similar features are
  also found in the LMC cluster NGC 1806 and the SMC cluster NGC
  411.}\label{F6}
\end{figure}

\begin{figure}[htbp]
\centering
\includegraphics[width=5.5in]{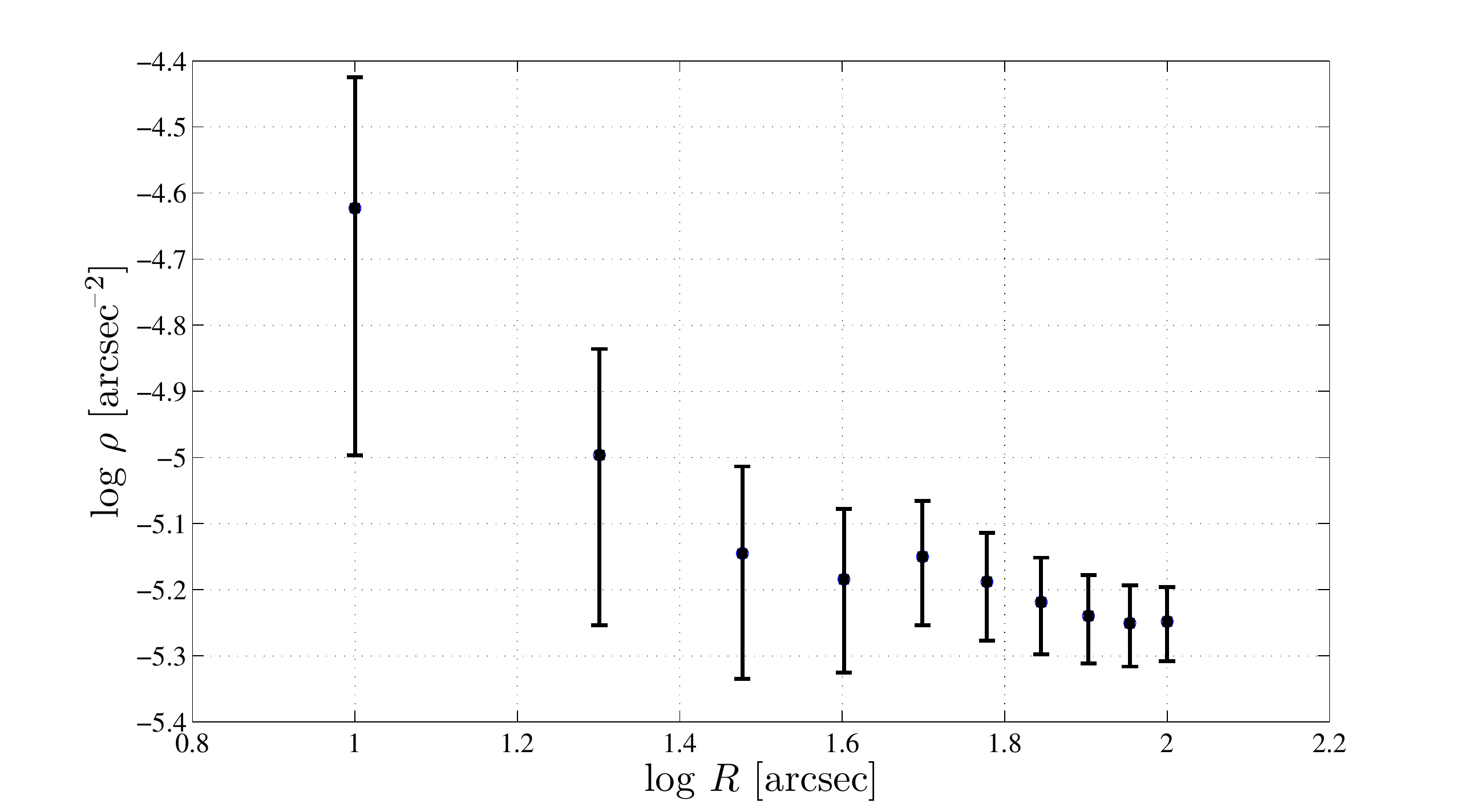}
\caption{Number-density profile of young stellar population stars in
  NGC 1783 (see also the CMD in Fig. \ref{F6}).}\label{F7}
\end{figure}

An unexpected discovery that relatively young, 1--2 Gyr-old star
clusters can harbor secondary stellar populations was made by
\cite{Li16a}, who found that the LMC clusters NGC 1783 and NGC 1806,
as well as the SMC cluster NGC 411, host secondary stellar populations
that are genuinely younger than their first-generation stars by about
500 Myr to 1 Gyr. The observed, different young stellar populations
are tightly associated with younger isochrones, indicating that they
are strictly coeval. However, these young stellar populations are
somewhat less centrally concentrated compared with the
first-generation RGB stars in the same clusters. These authors
speculated that the younger stars may have an external origin, which
renders the gas-accretion scenario promising once again. However,
because the observed young stellar populations in those extragalactic
star clusters only occupy minor mass fractions, additional numerical
simulations are required to further study the implications of these
observations. In Fig. \ref{F6} we present the observed CMD of NGC 1783
as an example. Although the young stars have a slightly more extended
nature compared to the approximately equal-luminosity RGB stars, they
are firmly concentrated within 1 to 2 core radii. The observed young
stellar population stars have a number-density profile that clearly
peaks in the clusters' core regions. This cannot be explained by
invoking background contamination (see Fig. \ref{F7}), which implies
that these newly born stellar populations indeed physically belong to
their host star clusters.

Another primordial formation scenario is based on cluster mergers,
which is the prevailing model for some of the most massive GCs, such
as NGC 1851 \citep{Carr10a}. An advantage of this hierarchical
formation scenario is that it solves the `initial exhaustion'
problem. This scenario preferentially favors clusters that are located
in crowded environments, such as the so-called `clusters of clusters'
\citep[e.g., cluster pairs in the Antennae interacting
  system;][]{Bast09b}. Cluster members in these environments are
expected to frequently collide and merge \citep{Amar13a}. However,
most GCs in our Milky Way are located in the Galactic halo, where
frequent mergers are not expected to occur.

The evolutionary scenario attributes the observed abundance variations
to the dredge-up of material that has been processed through the CNO
cycle in the cluster stars themselves. Because the observed chemical
dispersions of different elements in GC stars only represent
variations in the stellar {\it surface} abundances, any process that
can somehow transfer core material to the stellar surface would
produce star-by-star variations in elemental abundances. Because the
evolutionary scenario strongly depends on the efficiency of stellar
convective mixing, various mechanisms that may affect the mixing of
stellar material have been proposed. The most promising scenario
suggests that stars will undergo deep mixing when they evolve off the
MS. \cite{Lee10a} carefully studied the Na--O distributions in
Galactic GCs. They found a dependence of the Na--O anticorrelation on
RGB luminosity. This is probably owing to the internal deep mixing of
evolved stars during their ascent of the RGB. Their results thus
partially support the evolutionary scenario.

Another possible mechanism considers stellar magnetic fields:
\cite{Nucc14a} proposed a scenario involving the advection of
thermonuclear ashes by magnetized domains emerging near the H shell to
explain AGB-star abundances. Based on magnetohydrodynamics
calculations, they verified that the stellar-envelope crossing times
are sufficient to facilitate chemical dispersion in a huge convective
shell. They claimed that magnetic advection is a promising mechanism
for deep mixing, which may explain the observed abundance anomalies in
GCs. However, because stars in GCs are usually late-G- or K-type
stars, the central temperatures of individual low-mass stars do not
reach the threshold for producing the observed Na--O
anticorrelation. \cite{Jiang14a} proposed a model based
  on fast rotators that are produced by binary mergers or
  interactions. They found for binary merger products which are
  normal, rapidly rotating stars earlier than F-type, that these
  rapidly rotating stars can produce a Na--O anticorrelation with high
  significance. They claimed that binary interactions are a possible
solution to the chemical anomalies in GCs.

\begin{figure}[htbp]
\centering
\includegraphics[width=5.0in]{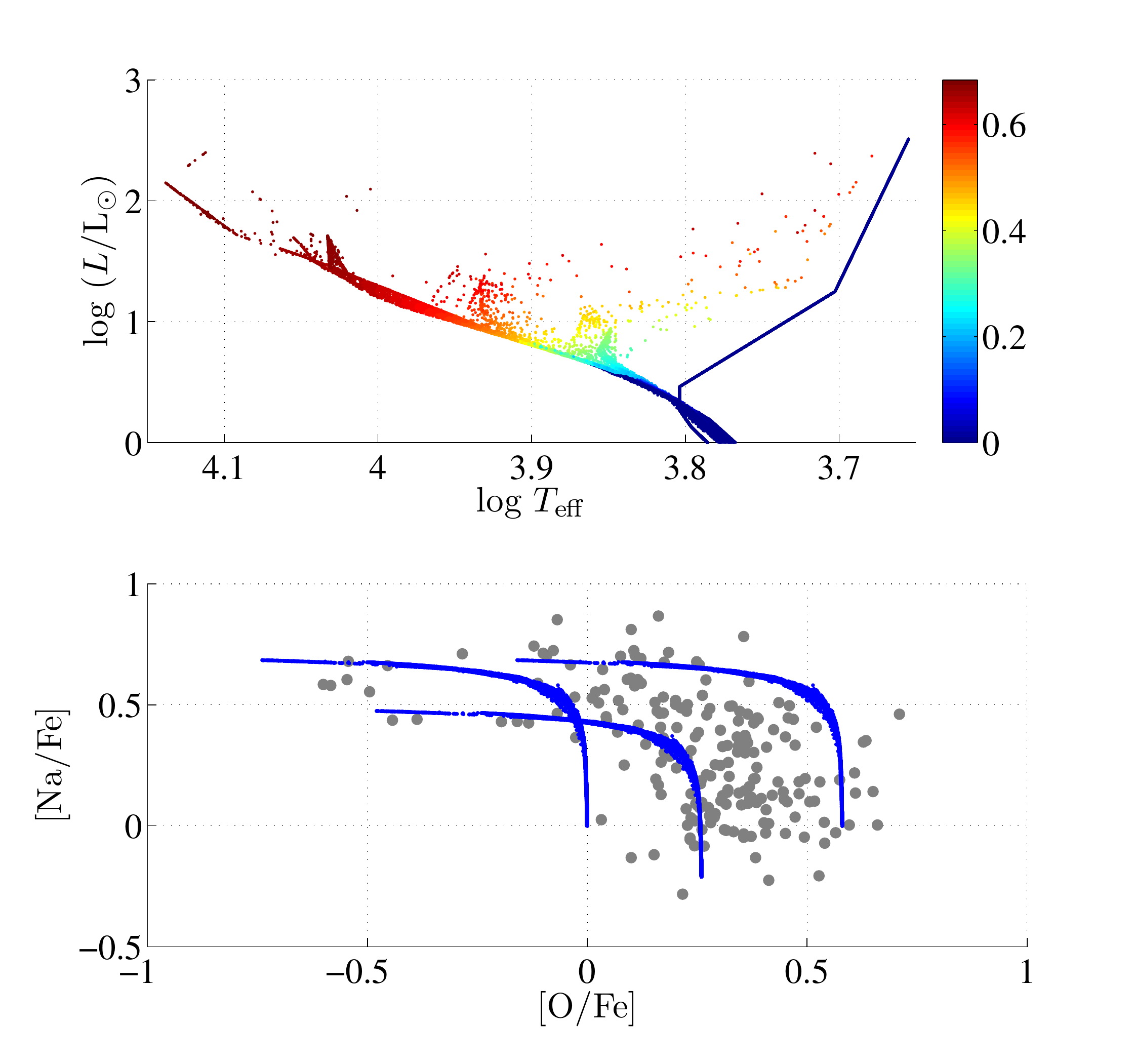}
\caption{The importance of binary systems in realistic stellar
  populations. Top: CMD of an SSP including binary interactions. Color
  bar: stellar Na abundance. Bottom: Na--O relationships produced by
  the model (blue) compared with observations (gray bullets). Each
  sequence of blue dots represents an SSP, characterized by different
  initial abundances.}\label{F8}
\end{figure}

The evolutionary scenarios do not require an extremely massive origin
for clusters to generate abundance anomalies. \cite{Jiang14a}'s
scenario only depends on a cluster's binary fraction. This implies
that OCs would also display Na--O anticorrelations if they are
sufficiently old for large numbers of their binaries to have
merged. However, so far we do not have any conclusive evidence that
OCs may harbor multiple stellar populations. \cite{Cant14a} studied
the chemical homogeneity of RC stars in the OC NGC 6751, but no
obvious correlations or anticorrelations among Al, Mg, Si, and Na were
found. A similar conclusion was also reached by \cite{Brag14a}, who
studied 35 evolved stars in an old OC, NGC 6791 \citep[8--9 Gyr
  old;][]{Brog12a}. They did not detect any significant star-to-star
chemical dispersions in C, N, O, or Na either. In Fig. \ref{F8} we
present the performance of the scenario of \cite{Jiang14a}. In the top
panel, we show the theoretical CMD of an SSP following binary mergers;
in the bottom panel, we show the Na--O anticorrelation produced by
this model compared with observations made in GCs.

Figure \ref{F8} shows that there is a clear correlation between the
stellar Na (as well as the O) abundance and luminosity. The binary
products are, in fact, `blue straggler stars' (BSSs). For more details
about the primordial and evolutionary scenarios for the origin of
multiple stellar populations in GCs, we recommend the review by
\cite{Kraf94a}; see also \cite{Deni15a}.

\section{Prospects}\label{S5}
\subsection{Studying Stellar Populations in Supermassive Star Cluster Candidates using Next-Generation Telescopes}

The origin of multiple stellar populations in star clusters is a
subject of significant current debate. To reach closure on this
question, the most straightforward approach would consist of observing
a `young GC' and resolving its initial stellar population(s). Most
scenarios suggest that a cluster's mass plays a key role in the
existence of multiple populations. The most popular scenarios suggest
that GCs should initially have been 10 to 100 times more massive than
their current masses. Therefore, a `good' candidate for follow-up
observations should be a young cluster with a mass of {\it at least}
$10^6$ $M_{\odot}$. To date, no such supermassive young clusters have
been observed in either the Milky Way or its satellites.

Strictly speaking, young GCs should reside at a redshift of $z \sim
2$, which is even beyond the ability of the {\sl James Webb Space
  Telescope} to resolve. Therefore, resolving such young clusters will
not be feasible in the near future. However, some candidate extremely
massive young GCs have already been detected in nearby starburst
galaxies. \cite{Smit06b} found five possible supermassive cluster
candidates in the galaxy M82. The mass of one of these `clusters,'
M82-A1, is $1.3^{+0.5}_{-0.4}$ $\times$ $10^6$ $M_{\odot}$, at an age
of only $6.4 \pm 0.5$ Myr. If this object truly is an individual YMC,
it may be the best candidate GC predecessor known. However, our
current facilities are not yet able to resolve individual stars in
such supermassive young star clusters, and only synthetic spectra are
available to study their stellar populations \citep[e.g.,][]{Cabr16a}.

Once next-generation telescopes have been developed, a top priority to
improve our understanding in this field will be to resolve individual
stars in these candidate young GCs. If these supermassive young
objects are confirmed as individual star clusters (and not blended
pairs or groups of clusters), then studying their stellar populations
will significantly help us constrain the origin of GCs. Using the CMDs
of these supermassive young star clusters, we can determine their ages
and any internal age dispersions, which is of fundamental importance
for studying the star-formation mode in young GCs. If the multiple
stellar populations observed in GCs represent multiple episodes of
star formation, then one should detect dispersed age distributions in
all supermassive young star clusters. This means that large fractions
of pre-MS stars are expected to reside in those clusters, and their
residual gas should still represent a large mass fraction.

\subsection{Measuring Stellar Rotation Rates in Intermediate-age Star Clusters}

The `eMSTO problem' associated with intermediate-age star clusters has
shown that rapid stellar rotation is an important process that may
confuse our understanding of the SFHs of star clusters. Although
measuring stellar rotation rates of individual stars in compact star
clusters at the distance of the LMC is challenging, some current
instruments can already obtain direct measurements of stellar rotation
in LMC star clusters. Using the Multi Unit Spectroscopic Explorer
(MUSE) at the European Southern Observatory's Very Large Telescope
(VLT), one can in principle obtain low-resolution spectra of all stars
resolvable by MUSE in a 1 arcmin$^2$ field covering the central
regions of intermediate-age LMC star clusters, including of their
F-type eMSTO stars. If rotation is the cause of the eMSTO, one will
immediately obtain a clear and unequivocal signature in support of
this.

\begin{figure}[htbp]
\centering
\vspace{-1cm}
\includegraphics[width=5.0in]{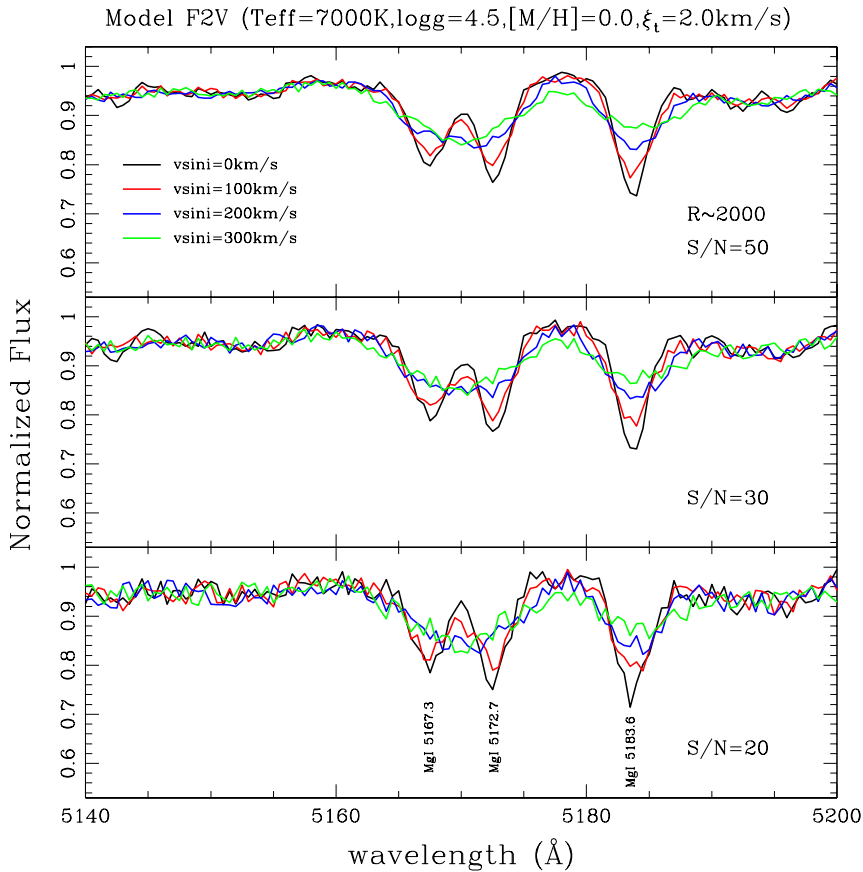}
\vspace{-3cm}
\caption{Simulations of the Mg{\sc i} triplet of an F2V star for
  different rotation velocities of 0 km s$^{-1}$ (black), 100 km
  s$^{-1}$ (red), 200 km s$^{-1}$ (blue), and 300 (green) km
  s$^{-1}$. The rotation axis is oriented perpendicularly to the line
  of sight. From top to bottom, the signal-to-noise ratio (S/N) = 50,
  30, 20; the adopted spectral resolution is $R \sim 2000$. If the
  mechanism responsible for the eMSTOs in intermediate-age star
  clusters is rapid rotation, then it will be detected at the minimum
  S/N = 30 for most of the lines.}\label{F9}
\end{figure}

One can quantify stellar rotation rates by measuring line broadening
in eMSTO stars. The Mg{\sc i} triplet around 5175 {\AA} is an ideal
indicator of the stellar rotation rate. Adopting the rapid stellar
rotation scenario, a stellar rotation rate from zero to a maximum of
70\% of the break-up rate will be responsible for the observed eMSTO
in intermediate-age star clusters. This corresponds to equatorial
velocities of $\sim$300 km s$^{-1}$. The MUSE spectral resolution will
allow us to resolve differences between no rotation and rotation rates
as low as (50--)100 km s$^{-1}$, with a signal-to-noise ratio of
30. In Fig. \ref{F9} we present the simulated morphology of the Mg{\sc
  i} triplet of F2V stars for different rotation velocities, as they
would be observed with MUSE. As long as there are no nearby stars in
the same resolution element with intensities within 1 mag of the
target stars, one will be able to derive a sample of stellar radial
velocities, velocity dispersions, and the overall distribution of
(eMSTO and other) rotation rates in such clusters. Such a project will
provide a definitive answer to the importance of the stellar rotation
scenario with respect to the internal age-spread model as regards the
origin of eMSTOs in intermediate-age star clusters.

\subsection{Exploring the Relationship between Elemental Abundances and the Luminosities of Blue Straggler Stars}

If the scenario of \cite{Jiang14a} is correct, a correlation between
elemental abundances and the luminosities of BSSs is expected. BSSs
are rejuvenated MS stars, formed through either stellar collisions
\citep{Hills76a} or binary mass transfer \citep{McCr64a,Carn01a}. The
relative importance of these two formation channels during different
evolutionary stages in star clusters has not yet been determined
unequivocally from an observational perspective. However, numerical
simulations have shown that the number of BSSs of binary origin will
continue to dominate the total BSS sample over cosmic timescales
($\sim$10 Gyr). The number of collisional BSSs becomes comparable to
the number of BSSs originating from binary interactions on timescales
exceeding a Hubble time \citep{Hypk13a}.

\begin{figure}[htbp]
\centering
\includegraphics[width=5.0in]{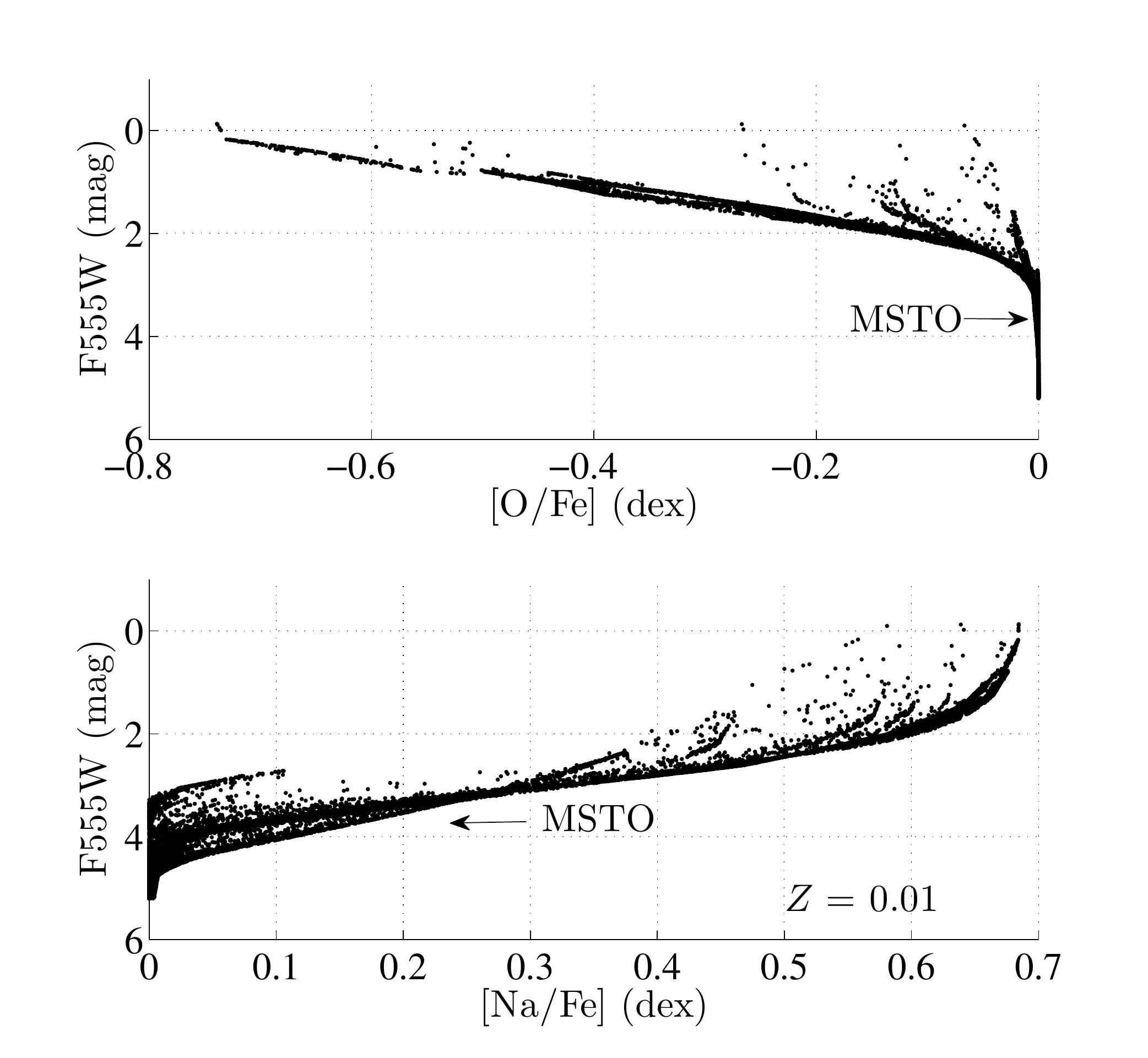}
\caption{Stellar F555W magnitude as a function of (top) [O/Fe] and
  (bottom) [Na/Fe] abundance, based on \cite{Jiang14a}'s model. The
  arrows indicate the typical magnitude of the MSTO region. The
  metallicity adopted is $Z = 0.01$.}\label{F10}
\end{figure}

\cite{Jiang14a}'s scenario draws on binary mergers to generate a
population of fast rotators, producing surface abundance anomalies
through evolution of the latter. A promising approach to examine this
scenario is by studying the correlation between luminosity and
elemental abundances as pertaining to BSSs. In Fig. \ref{F10} we
present the relationship between the expected stellar oxygen and
sodium abundances ([O/Fe], [Na/Fe]) and their absolute F555W-band
magnitudes, calculated using the PARSEC stellar evolutionary models
\citep{Bres12a} for a metallicity of $Z = 0.01$. A clear correlation
between luminosities and elemental abundances appears, especially for
stars that are brighter than the typical magnitude of the MSTO
region. To measure a significant number of BSS magnitudes and chemical
abundances, relatively old OCs would be good targets. Since their
member stars have already evolved to old ages, there should have been
sufficient opportunities for binary mergers to occur. An ideal target
is the 6.5 Gyr-old OC NGC 188 \citep{Dema92a}. \cite{Math09a} found
that approximately 76\% of its BSSs are members of binary systems,
indicating a large fraction of interacting binary stars.

\subsection{Studying Ambient Gas Accretion Through Numerical Simulations}

Significantly more work is required before we can connect young GCs to
genuine GCs (which are usually defined as more massive than $10^5$
M$_{\odot}$---see Fig. \ref{F1}---and older than 10 Gyr). Numerous
studies have addressed this issue. In general, YMCs should contain
sufficient numbers of member stars initially to survive for a Hubble
time \citep{Port10a}. Another key parameter that determines the
survival time of YMCs is the slope of its stellar initial mass
function \citep{grijs07a}. Current consensus on long-term cluster
survival implies that the combination of internal and external cluster
dynamics must play a important role \citep{grijs10a}.

The discovery of young stellar populations in intermediate-age LMC
star clusters is exciting. It indicates that star clusters may have
access to various channels to form secondary stellar populations
\citep{Li16a}. A key difference between the observed secondary stellar
populations in these young GCs compared with old stellar populations
in genuinely old GCs relates to their dynamics. The young stellar
populations in the young clusters are less centrally concentrated than
the evolved stars of the first generation. In old GCs the secondary
stellar generations are usually more densely distributed than the
first-generation stars. Therefore, it is still unclear if the observed
secondary stellar generations in these relatively young clusters are
similar to the second-generation stars in old GCs.

More evidence for more evolved star clusters (e.g., with ages between
3 Gyr and 10 Gyr) is indeed required. Unfortunately, there is an age
gap between 3 Gyr and 10 Gyr in the LMC cluster sample
\citep{Balb10a}, so one has to turn to numerical simulations to study
this aspect. \cite{Li16a} proposed that ambient gas accretion may be a
possible solution to the observed young stellar populations. They
speculated that GMCs are possible sources of accreted gas. If this is
correct, then accretion should be a common process for all massive
star clusters (irrespective of their ages) that are located in
gas-rich galactic disks. However, the nature of the slightly less
concentrated spatial distributions of the young stellar populations
compared with their first-generation RGB counterparts is unresolved.
\cite{Cabr16b} argued that the observed features may have been caused
by misunderstood issues in field-star decontamination, but it is
highly unlikely that field-star contamination could artificially
produce such tightly constrained, well-populated stellar populations
(see Fig. \ref{F6}), which also exhibit clearly centrally concentrated
radial profiles (see Fig. \ref{F7}). Additional studies of the
dynamics of these younger stellar populations are required: indeed,
one needs to combine $N$-body simulations with up-to-date insights
into stellar evolution to explore these issues. More accurate models
should also place clusters in an external gravitational field, which
will affect the `evaporation' processes in different stellar
populations \citep[for more details, see][]{Spit87a}. More detailed
calculations should also use smoothed-particle hydrodynamics
simulations to trace accretion flows between star clusters and GMCs.

\section{Summary}\label{S6}

In this review, we have introduced the community's up-to-date insights
into the physics governing stellar populations in star clusters. We
have shown that because of initial gas expulsion, most star clusters
are not expected to exhibit multiple episodes of star formation. This
renders the origin of the common observation of multiple stellar
populations in star clusters an intriguing open question. The
observational status of stellar populations in star clusters can be
summarized as follows:

\begin{itemize}
\item For star clusters younger than $\sim$100 Myr, no conclusive
  evidence exists that they may harbor multiple stellar populations or
  ongoing star formation; no residual gas has been detected in
  extremely young clusters \citep{Bast14a}.
\item The eMSTO morphologies of some YMCs with ages older than 100 Myr
  \citep[e.g., NGC 1850 and NGC 1856;][]{Milo15a,Bast16a} are
  inconsistent with the expectations from SSPs. It appears that such
  eMSTOs are a common feature of intermediate-age star clusters in the
  LMC and SMC. However, spectroscopic analyses have shown that their
  member stars do no exhibit abundance anomalies \citep{Mucc14a}.
\item The presence of multiple stellar populations in old GCs is
  irrefutable: both the morphologies of their photometric features and
  star-to-star chemical abundance variations challenge the SSP
  scenario.
\end{itemize}

Various scenarios have been proposed to explain these observed
deviations from genuine SSPs. For young and intermediate-age star
clusters, age spreads (which favor eSFHs) and rapid stellar rotation
(which suggests that clusters are SSPs) are in competition. For old
GCs, all scenarios can be classified as either primordial or
evolutionary.

We have proposed a number of projects that seem feasible in the near
future and which may shed light on our understanding of stellar
population problems in star clusters. These include direct
measurements of stellar rotation rates in compact star clusters,
studies of the elemental abundances of BSSs, and the use of numerical
simulations to study gas accretion. Since a number of possible
candidate young GCs have been identified in nearby starburst galaxies,
employing next-generation telescopes to study these objects will
significantly contribute to an improved understanding of the origin of
stellar populations in star clusters.

\begin{acknowledgements}
C. L. is partially supported by a Macquarie Research Fellowship and by
Strategic Priority Program `The Emergence of Cosmological Structures'
of the Chinese Academy of Sciences (grant XDB09000000). R. d. G. and
L. D. acknowledge research support from the National Natural Science
Foundation of China through grants 11073001, 11373010, and 11473037.
\end{acknowledgements}

\label{lastpage}

\end{document}